\documentclass[english,a4paper,twocolumn,amsmath,amssymb,tightenlines,reprint,superscriptaddress]{revtex4-2}
\usepackage{bm}
\usepackage{graphicx,color}
\usepackage{color,soul}
\usepackage[colorlinks,linkcolor=magenta,citecolor=magenta]{hyperref}
\usepackage{amsmath}
\usepackage{physics}
\usepackage{mlmodern}

\newcommand{\mean}[1]{\langle #1 \rangle}
\newcommand{\Int}[1]{\int\dd #1\;}
\newcommand{\IInt}[3]{\int_{#2}^{#3}\dd #1\;}
\renewcommand{\vec}[1]{\bm{\mathbf{#1}}}

\newcommand{\al}{\alpha}
\newcommand{\gam}{\gamma}
\newcommand{\Gam}{\Gamma}

\newcommand{\kap}{\kappa}
\newcommand{\lam}{\lambda}

\newcommand{\x}{\vec r}

\newcommand{\im}{\text{i}}

\begin{document}

\title{Statistics and morphologies of stable droplets in scalar active fluids}

\author{Kathrin Hertäg}
\affiliation{Institute for Theoretical Physics IV, University of Stuttgart, Heisenbergstr.\ 3, 70569 Stuttgart, Germany}
\author{Joshua F. Robinson}
\affiliation{STFC Hartree Centre, Sci-Tech Daresbury, Warrington, WA4 4AD, United Kingdom}
\affiliation{H.\ H.\ Wills Physics Laboratory, University of Bristol, Bristol BS8 1TL, United Kingdom}
\author{Thomas Speck}
\email{thomas.speck@itp4.uni-stuttgart.de}
\affiliation{Institute for Theoretical Physics IV, University of Stuttgart, Heisenbergstr.\ 3, 70569 Stuttgart, Germany}

\begin{abstract}
  Conventional phase segregation is controlled by a positive interfacial tension, which implies that the system relaxes towards a state in which the interfacial area (or length) is minimized, typically manifesting as a single droplet that grows with the system size. Intriguingly, the extension of the underlying Model B paradigm by two non-potential terms (Active Model B+) is able to describe the stable coexistence of many finite droplets. Here we numerical study Active Model B+ in the vicinity of the transition between a single droplet (macrophase segregation) and multiple droplets (microphase segregation). Our results show that, although noise shifts transitions, the overall agreement with the mean-field theoretical predictions is very good. We find a strong correlation of droplet properties with a single parameter that determines the number, density, and the fractal dimension of droplets. Deeper inside the droplet phase we observe another transition to a hexagonal lattice of regular droplets.
\end{abstract}

\maketitle


\section{Introduction}

The notion that nature exploits phase separation to compartmentalize and recruit or sequester specific proteins, and that this process is involved in cellular functions, has led to a flurry of activities at the interface of the life sciences, physics, and chemistry~\cite{banani17,alberti21,visser24,alberti25}. While phase separation of simple passive mixtures is well understood, segregation in crowded, viscoelastic, heterogeneous, and chemically driven environments, as indeed found in living cells, stretches the existing framework of statistical physics to its limits. Our theoretical understanding of phase separation in mixtures with many interacting components driven away from thermal equilibrium is still in its infancy.

One intriguing observation is that phase segregation of proteins in solution \emph{in vitro} leads to coarsening, i.e., the rapid formation of large dense domains that then slowly grow. The same protein \emph{in vivo}, however, might forgo coarsening and form finite droplets, so-called condensates or coacervates, the sizes of which are stable over an extended period of time (one example is the cofactor BRD4 involved in transcription~\cite{sabari18,han20}). Several microscopic explanations have been proposed for this behavior: First is the impact of the viscoelastic environment, which requires accounting for the energy that is required to deform the host material in order to accommodate the droplet~\cite{style18,tanaka22,liu23}. Second, due to the many species of biomolecules involved, some might effectively act as ``surfactants'' and reduce the interfacial tension, thus stabilizing finite-sized droplets~\cite{folkmann21,bauer24}. And third, models that incorporate chemically driven exchanges between droplet and surroundings also show the emergence of stable droplets~\cite{weber19,julicher24}. So far, however, the experimental evidence seems insufficient to discriminate these scenarios for condensates in living cells. Ultimately, it might well be that a combination of microscopic processes is responsible given the compositional and structural complexity of \emph{in vivo} condensates.

Experimental data such as microscopy images and spectroscopy data is taken on length and time scales that implies local averages over many molecules. Typically, one or two molecular species are labeled and monitored. Even though proteins might undergo transitions between different conformations that alter their interactions, their total number is (approximately) conserved over these scales. Such conservation laws profoundly constrain mixtures and enable effective evolution equations of the \emph{slow} conserved quantities~\cite{halatek18,brauns20}. Naturally, they take the form of Cahn-Hilliard-like equations but may include additional terms that cannot be absorbed into an effective free energy~\cite{robinson25}. For a single effective component, the limiting field theory has been introduced in the context of motile active matter and has been coined ``Active Model B+'' (AMB+)~\cite{tjhung18} in reference to the paradigmatic ``Model B'' describing the phase separation kinetics of a conserved order parameter that is decoupled from hydrodynamic flows~\cite{hohenberg77}. AMB+ has been argued to govern specific systems such as chemically active mixtures~\cite{alston22}, active chiral hard disks~\cite{kalz24}, and active Brownian particles~\cite{bickmann20,speck22a,vrugt23}. Beyond applications to active fluids and field theory~\cite{cates17,cates19}, extensions of AMB+ to underdamped dynamics~\cite{vrugt23}, multicomponent mixtures~\cite{chiu24}, and the addition of non-conservative terms~\cite{li21} have been discussed.

\begin{figure*}[t]
  \centering
  \includegraphics{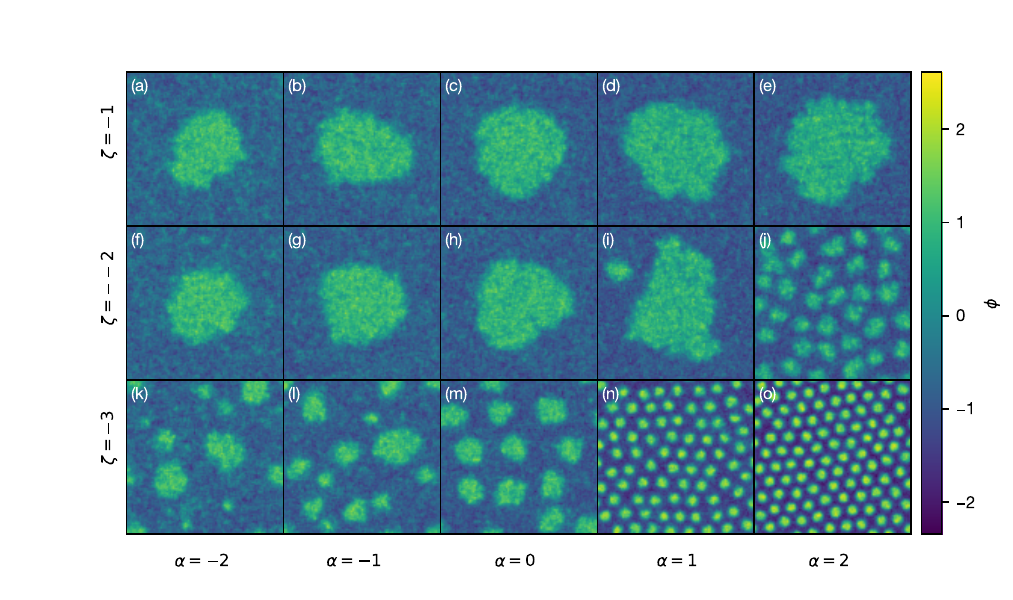}
  \caption{Steady-state simulation snapshots showing the field $\phi$ varying $\lam$ and $\zeta$ in Eq.~\eqref{eq:j} with $\al=(\zeta-2\lam)/\kap$. The global density is $\phi_0=-0.4$ and the noise strength is set to $D=0.3$. There are three regimes: (a-i)~macroscopic phase separation with a single dense domain, (j-m)~finite-sized droplets are stable, and (n,o) may order into a hexagonal lattice.}
  \label{fig:snap}
\end{figure*}

The most striking feature of AMB+ is that it exhibits a region in parameter space where finite-sized droplets are stable in steady state, at least qualitatively in agreement with observations for \emph{in vivo} protein condensates. While the microscopic details enter through the coefficients of the theory and their expressions have to be established for specific applications, AMB+ thus promises general insights into the physics underlying non-equilibrium microphase segregation such as negative effective interfacial tension leading to reverse Ostwald ripening~\cite{tjhung18}. Compared to the extensive literature on passive phase separation and coarsening, our knowledge of active \emph{microphase} separation in AMB+ is still limited~\cite{cates25}. To assess its scope and relevance as a minimal model for non-equilibrium phase segregation, in particular for the mentioned proteins condensates, requires a more detailed quantitative understanding of droplet properties. Recent studies have investigated capillary waves~\cite{fausti21}, the nucleation of droplets~\cite{cates23}, the distribution of droplet sizes~\cite{fausti24,yan25} (which can be mapped onto an effective ``bubble'' model of droplet radii), as well as coarsening kinetics~\cite{yadav25} and its connection to hyperuniformity~\cite{zheng24,deluca24}. The phase diagram of AMB+ has also been studied through renormalization approaches~\cite{caballero18,speck22a,papanikolaou24,fejos25}~\footnote{Although we note that the non-potential terms in Eq.~\eqref{eq:j} generate a cubic non-linearity in the effective free energy that ``runs off''~\cite{papanikolaou24}.}.

In this work, we further investigate the properties of steady-state droplets within AMB+. In Sec.~\ref{sec:theory}, we start by introducing the model and review the idea of an integrating factor to absorb non-potential interfacial terms into an effective bulk potential~\cite{solon18,solon18a,speck21b,speck25}. This exact mapping allows to use established tools to predict phase coexistence, at least on the mean-field level. We make conceptual progress through including the effect of finite droplets on the effective chemical potential, which allows to construct the coexisting densities without going through an evolution equation for the droplet radius. In Sec.~\ref{sec:sim}, we present the results of extensive numerical steady-state simulations with and without noise (representative snapshots are shown in Fig.~\ref{fig:snap}). We calculate a number of structural descriptors focusing on quantities that could be compared with experimental data extracted from microscopy images.


\section{Theory}
\label{sec:theory}

\subsection{The model}

We are interested in the evolution of a conserved field $\phi(\x,t)$. The continuity equation reads
\begin{equation}
  \partial_t\phi + \nabla\cdot\vec j = \eta
  \label{eq:phi}
\end{equation}
with deterministic current
\begin{equation}
  \frac{\vec j}{M} = -\nabla\left(\frac{\delta\mathcal{F}}{\delta\phi} + \lambda\abs{\nabla\phi}^2\right) + \zeta(\nabla^2\phi)\nabla\phi.
  \label{eq:j}
\end{equation}
Here, $\eta(\x,t)$ is a Gaussian white noise field conserving $\phi$ with mean $\mean{\eta(\x,t)}=0$ and correlations
\begin{equation}
  \mean{\eta(\x,t)\eta(\x',t')} = -2M D\nabla^2\delta(\x-\x')\delta(t-t'),
  \label{eq:noise}
\end{equation}
the strength of which is set by temperature $D$ (in natural energy units). A constant mobility $M$ can be absorbed into a redefinition of time, so without loss of generality we set $M = 1$. These dynamics conserve the global ``mass''
\begin{equation}
  \phi_0 = \frac{1}{A}\Int{^2\x}\phi(\x,t)
\end{equation}
in a system with area $A$. The particle current involves the derivative of the Ginzburg-Landau free energy functional
\begin{equation}
  \mathcal{F}[\phi] = \Int{^2\x}\left\{f(\phi)+\frac{\kappa}{2}\abs{\nabla\phi}^2\right\},
  \quad 
  f(\phi) = \frac{a}{2}\phi^2+\frac{u}{4}\phi^4
  \label{eq:gl}
\end{equation}
parametrized through the coefficients $a$, $u$, and $\kappa$, with the latter penalizing inhomogeneities of the field. Passive phase separation requires $a<0$ with the coexisting field values $\phi^\text{eq}_\pm=\pm\sqrt{-a/u}$ determined by the minima of the bulk free energy density $f(\phi)$. The two additional terms in Eq.~\eqref{eq:j} with coefficients $\lambda$ and $\zeta$ constitute terms that cannot be derived from a free energy and thus encode explicit breaking of detailed balance and time-reversal symmetry on the level of the field $\phi$~\footnote{But even the ``passive'' Model B might be derived from a microscopic model that breaks detailed balance, indicating that dissipation is associated with local degrees of freedom integrated out from the field $\phi$.}.

\subsection{Integrating factor}

We now summarize the formalism that allows us to extract the phase diagram and some properties of circular droplets in the noiseless limit closely following Ref.~\cite{tjhung18}. We employ polar coordinates and assume a symmetric and stationary droplet at the origin with profile $\phi(r)$. For such a profile not only the divergence $\nabla\cdot\vec j=0$, but also the radial component $j_r=0$ of the current, need to vanish. The $\zeta$-term in Eq.~\eqref{eq:j} is decomposed into a curl-free and a divergence-free part making use of the Helmholtz theorem, leading to $j_r=-\partial_r\mu$ with
\begin{equation}
  \mu = f'(\phi) - \kappa\nabla^2\phi - \frac{1}{2}(\zeta-2\lam)(\partial_r\phi)^2 + \mu_\zeta
  \label{eq:mu}
\end{equation}
taking the role of an effective chemical potential. Throughout the prime (applied to a function) denotes the derivative with respect to $\phi$. The last term in Eq.~\eqref{eq:mu} is a non-local term
\begin{equation}
  \mu_\zeta(r) = \zeta\IInt{r'}{r}{\infty} \frac{(\partial_{r'}\phi)^2}{r'}
  \label{eq:delta}
\end{equation}
due to the curved interface of the droplet. For a flat interface, this term vanishes. For a curved interface, $\mu_\zeta$ is non-zero (and approximately constant as long as $\nabla\phi\simeq0$) inside the dense droplet and drops through the interface to zero away from the droplet. In any case, in steady state $\mu=\bar\mu$ has to be constant and Eq.~\eqref{eq:mu} constitutes an ordinary differential equation of second order for the profile $\phi(r)$. To solve this equation, we need to supply boundary conditions as well as the value for $\bar\mu$.

To extract the coexisting bulk values $\phi_\pm$ without solving Eq.~\eqref{eq:mu} we look for a function $\rho(\phi)$ that allows us to transform the non-potential derivatives in Eq.~\eqref{eq:mu} into the total derivative of $\Psi$,
\begin{equation}
  \mu\partial_r\rho = -\partial_r\Psi + \chi,
  \label{eq:decomp}
\end{equation}
with non-potential remainder $\chi$. To this end, we make the ansatz
\begin{multline}
  \mu\partial_r\rho = \partial_r\left[\psi - \frac{1}{2}\kappa\rho'(\partial_r\phi)^2 + \rho\mu_\zeta \right] \\ = \left[\pdv{\psi}{\rho} - \kappa\partial_r^2\phi - \frac{1}{2}\kappa\frac{\rho''}{\rho'}(\partial_r\phi)^2 + \mu_\zeta \right]\partial_r\rho - \rho\zeta\frac{(\partial_r\phi)^2}{r}
\end{multline}
with point-wise function $\psi(\phi)$. The second line follows after differentiation. Comparing with Eq.~\eqref{eq:mu}, we read off
\begin{equation}
  \pdv{\psi}{\rho} = f'(\phi), \qquad
  \frac{\rho''}{\rho'} = \frac{\zeta - 2\lam}{\kappa} \equiv \al
  \label{eq:rho}
\end{equation}
so that the remainder in Eq.~\eqref{eq:decomp} becomes
\begin{equation}
  \chi = \frac{(\partial_r\phi)^2}{r}(\zeta\rho-\kappa\rho').
  \label{eq:chi}
\end{equation}
Here the second term originates from the Laplacian $\nabla^2\phi=\partial_r^2\phi+\tfrac{1}{r}\partial_r\phi$ expressed in polar coordinates in two dimensions. The condition Eq.~\eqref{eq:rho} is a differential equation that fixes
\begin{equation}
  \rho(\phi) = \frac{1}{\al}\left(e^{\al\phi}-1\right)
\end{equation}
depending on the single parameter $\al$ defined in Eq.~\eqref{eq:rho}. Integration constants have been chosen to recover $\rho\to\phi$ in the limit $\al\to0$, which includes passive Model B ($\zeta=\lam=0$).

\subsection{Effective action}

Through the integrating factor, we have determined the effective potential
\begin{equation}
  \Psi = \frac{1}{2}\kappa\rho'|\partial_r\phi|^2 - \psi - \rho\mu_\zeta.
\end{equation}
Exploiting $\mu=\bar\mu$ and rearranging Eq.~\eqref{eq:decomp} leads to $\partial_r(\Psi+\rho\bar\mu)=\partial_rP=\chi$ with another function
\begin{equation}
  P = \frac{1}{2}\kappa\rho'|\partial_r\phi|^2 - \rho\mu_\zeta + p, \qquad p \equiv \rho\bar\mu - \psi.
  \label{eq:P}
\end{equation}
Our notation already suggests that $\psi(\phi)$ takes the role of a bulk free energy while $p(\phi)$ corresponds to the bulk pressure, although we stress that this is a mapping onto an effective system and, e.g., $p$ does not equal the mechanical pressure. The total effective pressure $P$ includes contributions from the field gradient inside the interface.

In the language of analytical mechanics, $P$ would be called the Hamiltonian, which comes with the Lagrangian
\begin{equation}
  L = \frac{1}{2}\kappa\rho'|\nabla\phi|^2 + \rho\mu_\zeta - p
\end{equation}
obtained through a Legendre transform of the gradient. Indeed, a straightforward calculation shows that Eq.~\eqref{eq:mu} corresponds to the Euler-Lagrange equation
\begin{equation}
  \nabla\cdot\pdv{L}{(\nabla\phi)} - \pdv{L}{\phi} = 0.
\end{equation}
The Hamiltonian principle then assures us that the solution of the Euler-Lagrange equation yields the field $\phi(\x)$ for which the action functional
\begin{equation}
  \mathcal A[\phi] = \Int{^2\x} \left[\frac{1}{2}\kappa\rho'|\nabla\phi|^2 + \rho\mu_\zeta - p\right]
  \label{eq:action}
\end{equation}
is stationary. The comparison with Eq.~\eqref{eq:gl} thus strengthens the interpretation of $\psi$ as a bulk free energy with $p$ following from another Legendre transform for the conjugate variables $\mu$ and $\rho$. The use of such an action has not been discussed so far in connection with AMB+ and might allow progress through advanced sampling approaches that make use of variational methods.

\subsection{Coexistence}

\begin{figure}[t]
    \centering
    \includegraphics{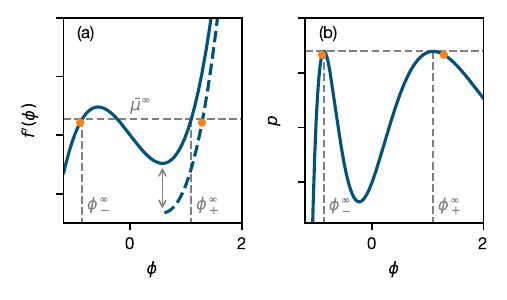}
    \caption{Coexistence in the mean-field theory. (a)~Effective bulk chemical potential $f'(\phi)$ and (b)~bulk pressure $p(\phi)$ as functions of field value $\phi$ for $\al=-2$. The gray vertical lines indicate the coexisting densities $\phi^\infty_\pm$ for a flat interface, which only depend on $\al$. For a droplet with radius $R$ the dense branch of $\mu$ (dashed line) is shifted by $\mu_\zeta(0)$ [Eq.~\eqref{eq:delta}], shown here at $\zeta=-4$ with radius $R=10$. The symbols show the coexisting field values for a droplet, for which $\phi_-$ is shifted to lower and $\phi_+$ to larger values, i.e., the droplet is more compact and the gas more dilute compared to a flat interface. Other parameters are $-a=u=0.25$ and $\kap=1$.}
    \label{fig:theory}
\end{figure}

For inhomogeneous systems, $\partial_rP=\chi$ now yields a second condition
\begin{multline}
  -\IInt{r}{0}{\infty}\chi = P(r\to 0) - P(r\to\infty) \\ = p(\phi_+) - (\rho\mu_\zeta)_{r=0} - p(\phi_-)
  \label{eq:coex:pre}
\end{multline}
in addition to the uniformity of $\mu=\bar\mu$. Here we have assumed that the integration limits correspond to the bulk phases with $\phi_\pm$ for which the gradient term in Eq.~\eqref{eq:P} vanishes. For a flat interface, $\chi=\mu_\zeta=0$ become zero so that Eq.~\eqref{eq:coex:pre} reduces to the customary equality of the effective pressure, $p(\phi^\infty_+)=p(\phi^\infty_-)$, with coexisting bulk densities $\phi^\infty_\pm$. Figure~\ref{fig:theory}(a) shows the effective bulk chemical potential $f'(\phi)$, which depends on $\zeta$ and $\lam$ only through their combination $\al$. At coexistence it is equal to $\bar\mu^\infty$, for which the effective bulk pressure is also equal with zero-slope tangent, cf. Fig.~\ref{fig:theory}(b). In contrast to passive systems, at $\zeta\neq 0$ a droplet with finite radius $R$ changes the effective chemical potential inside the droplet by $\mu_\zeta(0)$ [Eq.~\eqref{eq:delta}] while $\mu_\zeta$ is still zero for the gas away from the droplet. This shifts the chemical potential $\bar\mu(R)$ at coexistence and also the values $\phi_\pm(R)$ inside the droplet and in the surrounding gas as indicated by the symbols in Fig.~\ref{fig:theory}.

\begin{figure}[t]
  \centering
  \includegraphics{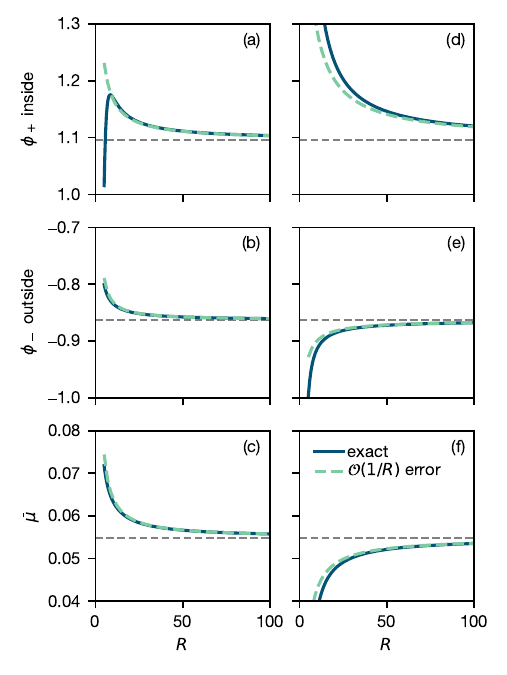}
  \caption{Coexisting densities $\phi_\pm(R)$ and effective chemical potential $\bar\mu(R)$ in mean-field steady state as a function of droplet radius $R$ for the same $\alpha=-2$: (a-c)~$\lam=0.5$, $\zeta=-1$ with $\gam>0$ and (d-e)~$\lam=-1$, $\zeta=-4$ with $\gam<0$. The approximate (dashed) and quasi-exact (solid) theories are described in the text. Vertical dashed lines indicate the limiting values for the flat interface. Other parameters are $-a=u=0.25$ and $\kap=1$.}
  \label{fig:theo}
\end{figure}

Figure~\ref{fig:theo} shows how $\phi_\pm(R)$ change with droplet size. To this end, we solve the weak formulation of $\nabla\cdot\vec j=0$ with polar symmetry in the domain $r \in [0, L]$ with boundary conditions $\phi'(0) = \phi'(L) = \phi(R) = 0$ via Galerkin's method of weighted residuals. We set the domain size $L = 5R$ so that gradients in $\phi$ are negligible approaching the outer boundary and discretize the domain into $N=101$ nodes representing $\phi(r)$ using 4th-order Hermite interpolating polynomials as weight functions at each node. In principle, this approach has no errors due to curvature, but will still have errors due to the discretization.

To make analytic progress in the regime of circular droplets with large but finite radius $R$, we expand the coexisting densities $\phi_\pm(R)\approx\phi_\pm^\infty+\beta_\pm/R$. Rearranging Eq.~\eqref{eq:coex:pre}, to the lowest order of $1/R$ the condition on the pressure becomes~\cite{tjhung18}
\begin{equation}
  p(\phi_+) - p(\phi_-) \approx \frac{\gam}{R}
  + \order{\frac{1}{R^2}}.
  \label{eq:coex}
\end{equation}
In this limit the coefficient
\begin{equation}
  \gam \equiv \IInt{x}{-\infty}{\infty} \kap_\text{eff}(\phi^\infty)(\partial_x\phi^\infty)^2
  \label{eq:gam}
\end{equation}
involves the solution $\phi^\infty(x)$ for the flat interface ($R\to\infty$) with $x$ the coordinate along the interface normal. Here we have assumed that the interfacial width is much smaller than $R$ and thus the change of the profile is concentrated in a small interval containing $R$, $\Int{r}r^{-1}f(r)\approx R^{-1}\Int{r} f(r)$. The same argument leads to $\mu_\zeta(0)=\zeta\Gam/R$ with $\Gam\equiv\IInt{x}{-\infty}{\infty} (\partial_x\phi^\infty)^2$ and we read off the point-wise function
\begin{equation}
  \kap_\text{eff}(\phi) \equiv \zeta\rho(\phi_+) - \zeta\rho(\phi) + \kappa\rho'(\phi)
\end{equation}
entering Eq.~\eqref{eq:gam}. In the passive limit ($\zeta=\lam=0$) we recover $\kap_\text{eff}\to\kap$ as expected, and Eq.~\eqref{eq:gam} is the well-known expressions for the interfacial tension~\cite{evans79}. Strikingly, through $\kap_\text{eff}$ the coefficient $\gam$ can become negative at variance with its behavior in passive systems, where $\gam>0$ is strictly positive. Finally, we expand $\bar\mu(R)\approx\bar\mu^\infty+\beta_\mu/R$, which we insert into the bulk pressure [Eq.~\eqref{eq:P}] to obtain $p(\phi_\pm)\approx p(\phi^\infty_\pm)+\rho(\phi^\infty_\pm)\beta_\mu/R$ using that $p'(\phi^\infty_\pm)=0$. Plugging this expansion into Eq.~\eqref{eq:coex}, we thus find the leading-order coefficient
\begin{equation}
  \beta_\mu = \frac{\gam}{\rho(\phi_+^\infty)-\rho(\phi_-^\infty)}
\end{equation}
governing the shift of $\bar\mu$ away from $\bar\mu^\infty$ for the flat interface. For $\gam<0$ clearly $\beta_\mu<0$ and thus $\bar\mu(R)$ becomes smaller for smaller droplets.

For $\al\neq 0$ we determine $\phi^\infty(x)$ numerically, see Appendix~\ref{appendix:flat} for details. In Fig.~\ref{fig:theo} we show the results of this large-$R$ approximation, which compare favorably with the quasi-exact numerical solution. Deviations only become significant at large curvatures as $1/R\sim 10$. With $\gamma > 0$ [Fig.~\ref{fig:theo}(a-c)] the exact curve shows that the droplet density $\phi_+$ begins to rapidly decay towards $\phi_-$ at extreme curvatures when $R$ approaches the interfacial width. This high-curvature effect is not captured by the approximate theory, and curiously seems to be absent in $\gamma < 0$ [Fig.~\ref{fig:theo}(d-f)] for $R \ge 5$. We encountered issues with numerical convergence for $R \lesssim 5$ and thus could not probe smaller droplets.

\subsection{Steady-state behavior}
\label{sec:steady}

Somewhat counter-intuitively, we can determine a current-free steady state for \emph{any} radius $R$ with coexisting densities $\phi_\pm(R)$ that simultaneously obey $f'(\phi_+)+\mu_\zeta(0)=f'(\phi_-)$ together with Eq.~\eqref{eq:coex}. So what is the radius $R_\ast$ that the droplet will adopt eventually? For $\gam>0$ we find the conventional scenario: Minimizing the interfacial contribution implies a circular droplet (on average), which reduces $\bar\mu$ [and thus the effective action Eq.~\eqref{eq:action}] through growth. In a finite system this growth is restricted by the conservation of the total ``mass'', which determines the final radius $R_\ast$. Minimization of the interfacial size under this constraint also implies that the system undergoes ``finite-size transitions'' between different morphologies as the conserved global density $\phi_0$ is varied~\cite{binder12}.

Intuitively, a negative effective tension $\gam<0$ implies a proliferation of interfaces and one might expect that the system becomes homogeneous again. Instead, simulations indicate that droplets with finite sizes become stable, and the steady state is a ``gas'' of droplets~\cite{tjhung18,fausti24}. To understand how this state emerges within the theoretical framework presented in this section, let us first consider two neighboring droplets with different radii $R_1>R_2$. Right outside each droplet, the effective chemical potentials are $\mu(R_1)>\mu(R_2)$. Hence, there is a gradient in $\mu$ driving a current from the big to the small droplet until both have the same size and $\mu$ is uniform. For a fixed number of droplets $n$ in a fixed area $A$, the conservation of mass expressed through the lever rule 
\begin{equation}
  \frac{\phi_0-\phi_-}{\phi_+-\phi_-} = \frac{n\pi\mathcal R^2}{A}
  \label{eq:lever}
\end{equation}
constrains the coexisting densities $\phi_\pm(R)$. The right-hand side of Eq.~\eqref{eq:lever} is the area fraction occupied by droplets. Note that so far $R$ has only entered as a parameter through Eq.~\eqref{eq:coex} but does not necessarily correspond to the ``true'' radius of droplets in the sense of the ``equimolar'' radius $\mathcal R$ of droplets obtained through replacing the radial profile through a sharp interface jumping from $\phi_+$ to $\phi_-$ at $\mathcal R$ (also known as ``Gibbs dividing surface''). Nevertheless, we will assume $\mathcal R\simeq R$. The lever rule now determines the target radius $R_\ast$, and droplet sizes (at fixed $n$) evolve towards this value until $\bar\mu(R_\ast)$ has become uniform. It does not, however, determine $n$.

In fact, droplets are inherently unstable since reducing their size (e.g. through splitting) reduces the effective chemical potential [Fig.~\ref{fig:theo}(f)]. At the same time, the background density $\phi_-$ drops [Fig.~\ref{fig:theo}(e)] and ``mass'' has to be moved into droplets. While droplets compactify as $R$ becomes smaller [Fig.~\ref{fig:theo}(d)], fluctuations might also initiate the nucleation of a new droplet with a radius much smaller than $R_\ast$. Since its local chemical potential is smaller than $\bar\mu(R_\ast)$, it is a sink for the other droplets, which will shrink (therefore further reducing the background $\phi_-$), whereas the new droplet will grow. Hence, $n$ increases by one and $R_\ast$ decreases in response to adding a new droplet. A dynamic steady state is reached when the rates for the coalescence of droplets (reducing $n$) and the nucleation of new droplets balance~\cite{fausti24}.


\section{Simulations}
\label{sec:sim}

\subsection{Simulation details}

To numerically solve Eq.~\eqref{eq:phi}, the field $\phi$ is spatially discretized on a regular grid with value $\phi_i(t)$ at lattice site $i$. In particular, the noise correlations Eq.~\eqref{eq:noise} then become
\begin{equation}
  \mean{\eta_i(t)\eta_j(t')} = 2D\ell^{-2}\Lambda_{ij}\delta(t-t')
\end{equation}
with lattice spacing $\ell$ and the graph Laplacian $\Lambda_{ij}=\delta_{ij}\sum_kA_{ik}-A_{ij}$ ($A_{ij}$ is one if sites $i$ and $j$ are neighbors and zero otherwise)~\cite{cavagna24}. We implement the noise by first generating a random current and then taking the divergence (using finite difference operators) to conserve $\phi$. To propagate the resulting finite-difference equations in two dimensions, we use a GPU-accelerated implementation~\cite{ambplusGitHub} of the same Euler forward discretization scheme as in Ref.~\cite{tjhung18}. Their $\order{\ell^2}$ Euler forward scheme uses a mixture of $\order{\ell^8}$ line and $\order{\ell^2}$ block stencils that were chosen to make integration numerically stable with large time step $\Delta t=0.01$. The stencils applied to different terms and the bespoke block stencils are given in Appendix~\ref{appendix:stencils}. Throughout this section, we choose as lattice spacing $\ell=\sqrt{\kap}$ so that $\kap$ becomes unity in these units and we fix $u=-a=0.25$ (same as in Ref.~\cite{tjhung18}) and vary $\lambda$ and $\zeta$.

\subsection{Noiseless simulations}

\begin{figure}[t]
  \centering
  \includegraphics{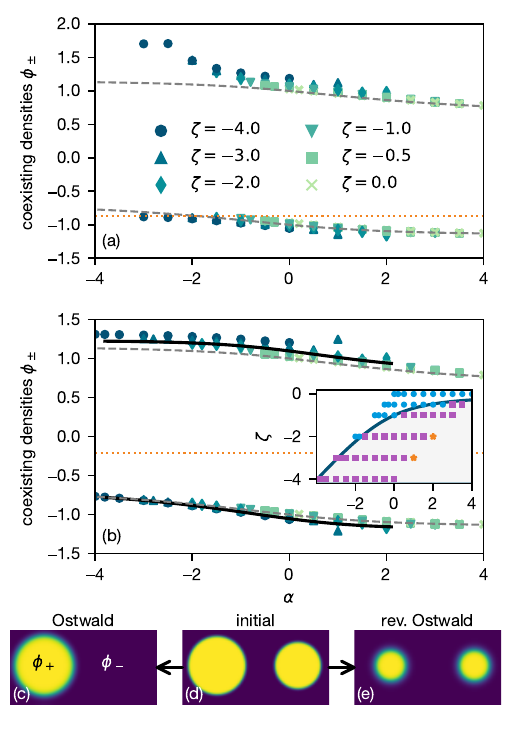}
  \caption{Coexisting densities $\phi_\pm$ as a function of $\alpha$ (the binodal) determined from noiseless ($D = 0$) simulations. Numerical results (symbols) starting (a)~from a single droplet and (b)~from two droplets. Dashed gray lines show $\phi_\pm^\infty$. The dotted lines indicate the global density $\phi_0$. The solid line in (b) is the self-consistent solution for $\zeta=-4$ constrained by the lever rule [Eq.~\eqref{eq:lever}] with $\mathcal R=R$. Inset: the number of final droplets in the $\al$-$\zeta$ plane is either a single droplet ($\bullet$), two droplets ($\blacksquare$), or more than two droplets ($\bigstar$). The solid line is the mean-field prediction at which $\gam$ changes sign, with negative $\gam<0$ inside the shaded area. The bottom row illustrates the protocol: (d)~initially two droplets with different radii are placed ($R_1=25$ and $R_2=20$), from which either (c)~the big droplet grows and the smaller one vanishes (Ostwald ripening) or (e)~both droplets approach the same size (reverse Ostwald ripening).}
  \label{fig:phi-alpha}
\end{figure}

We first perform noiseless simulations ($D=0$) with specific initial conditions to test our implementation and some of the predictions of the mean-field theory. We start with a single droplet (on a $128\times128$ grid) with initial condition
\begin{equation}
    \phi_i = \tanh(R_0-|\x_i-\x_0|),
    \label{eq:drop_field}
\end{equation}
where $R_0$ is the radius and $\x_0$ the center site of the droplet so that initially $\phi_\pm(0)=\pm 1$. We then advance the field until a steady state is reached and we determine $\phi_\pm$ and the final radius of the droplet. In Fig.~\ref{fig:phi-alpha}(a), the numerical results are plotted together with the theoretical mean-field prediction $\phi_\pm^\infty$ for a flat interface. While the overall agreement is very good despite the finite curvature of droplets, we observe deviations mostly for $\phi_+$ for negative values of $\al$. These deviations are explained easily since the initial state with $R_0=18$ implies $\phi_0\simeq-0.87$, which crosses the binodal as $\al$ becomes smaller and the system returns to the homogeneous state. Approaching the binodal, droplets become very small, further amplifying the impact of curvature. For most simulations, the final state remains a single droplet, although for strongly negative $\lam$ the droplet starts to deform away from the circular shape (resembling a square with rounded edges) and in two simulations splits into four droplets ($\lam=-2$ with $\zeta\in\{-2,-3\}$).

\begin{figure*}[t]
  \centering
  \includegraphics{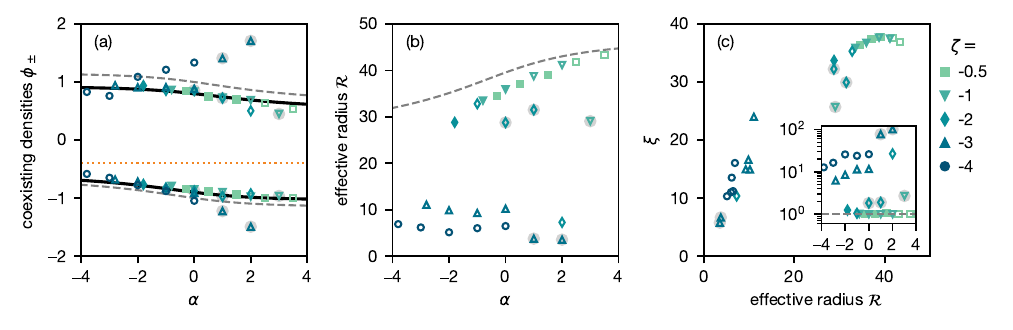}
  \caption{Characterizing droplets at global density $\phi_0=-0.4$ and non-zero noise strength $D = 0.3$ on an $128\times 128$ grid. (a)~The coexisting field values $\phi_\pm$ as a function of $\al$ indicating the different values of $\zeta$. The gray dashed lines are the theoretical mean-field predictions $\phi^\infty_\pm$, and the solid black lines have been obtained through multiplying $\phi^\infty_\pm$ with a constant factor (as a guide to the eye). (b)~Measured average droplet radii $\mathcal R$ as a function of $\al$. The gray dashed line corresponds to inverting the lever rule [Eq.~\eqref{eq:lever}]. (c)~The length scale $\xi$ extracted from the structure factor plotted as a function of the droplet radius. The dashed line indicates a single droplet. Inset: The average number of droplets as a function of $\al$. In all panels, closed symbols correspond to predicted $\gam>0$ and open symbols to $\gam<0$. Highlighted symbols are discussed in the main text.}
  \label{fig:coex}
\end{figure*}

In a second step, we place two droplets onto an elongated $128\times64$ grid using the same profile as in Eq.~\eqref{eq:drop_field} twice [Fig.~\ref{fig:phi-alpha}(d)]. We observe mass transport between droplets as predicted, either through forward [Fig.~\ref{fig:phi-alpha}(c)] or reverse Ostwald ripening [Fig.~\ref{fig:phi-alpha}(e)] depending on the values of $\zeta$ and $\lam$. The final values $\phi_\pm$ are plotted in Fig.~\ref{fig:phi-alpha}(b) together with the theoretical mean-field prediction $\phi_\pm^\infty$ for a flat interface, which are in good agreement. We also show the mean-field solution for the coexisting densities $\phi_\pm(R_\ast)$ for $\zeta=-4$ exploiting the lever rule [Eq.~\eqref{eq:lever}] to obtain the droplet radius $R_\ast$. Although taking into account curvature effects improves the agreement with the numerical results, fully quantitative agreement is not reached. We attribute the remaining differences to the $1/R$-approximation in Eq.~\eqref{eq:coex} and the assumption that $\mathcal R=R$ in Eq.~\eqref{eq:lever}.

The inset in Fig.~\ref{fig:phi-alpha}(b) shows the final number of droplets and thus allows to discriminate between forward and reverse Ostwald ripening of the initial two droplets. We find excellent agreement with the mean-field prediction. Venturing further into the regime of reverse Ostwald ripening, we observe the splitting of the two initial droplets into multiple droplets.


\subsection{Free droplets}

We now turn on the noise and perform simulations starting from a uniform random initial state and let the system relax into the steady state before we collect data. The global average value $\phi_0=-0.4$ is sufficiently far from the binodals to suggest spinodal decomposition and we indeed observe that droplets start to form quickly. Typical steady-state snapshots for the different parameters are shown in Fig.~\ref{fig:snap}.

The results are collected in Fig.~\ref{fig:coex}. We first turn to the density inside and outside of droplets. To this end, we construct histograms of the field values from the numerically sampled fields. These histograms exhibit two peaks as expected, and we use the peak positions to estimate $\phi_\pm$. In Fig.~\ref{fig:coex}(a), we plot $\phi_\pm$ as a function of $\al$. We find two qualitatively different regimes. The first regime corresponds to conventional phase separation with a single droplet at steady state. We observe that the coexisting densities $\phi_\pm$ collapse onto two master curves independent (at least within statistical uncertainties) of $\zeta$. These master curves are not exactly given by $\phi^\infty_\pm$ but seem to be shifted inwards by an approximately constant factor. We already note that some of the state points for which mean-field predicts $\gam<0$ still coarsen into a single droplet.

We perform an image analysis to isolate and analyze droplets directly. This is done in several steps: First, we employ Otsu's thresholding method (as implemented in \texttt{scikit-image}) to filter background noise and to assign droplet and background pixels. In this binary picture of black (background) and white pixels, single black pixels surrounded by white are filled and we join connected white pixels into clusters that we identify with droplets. Aggregates of less than five pixels are excluded. This way we obtain the size of each droplet in pixels and calculate their effective radius (which is used as an estimate for $\mathcal R$) assuming circular droplets.

Results for the effective radius $\mathcal R$ are shown in Fig.~\ref{fig:coex}(b) as a function of $\al$ for different $\zeta<0$. Fully phase-separated systems again collapse onto a single curve that is below the prediction of the mean-field theory, which we obtain through inverting the lever rule [Eq.~\eqref{eq:lever}]. There is a notable group of outliers for $\zeta=-1$ and $\zeta=-2$ falling below the master curve. Inspecting movies reveals that at these state points the droplet splits and merges over the course of the simulation so that sometimes two droplets are detected [cf. Fig.~\ref{fig:snap}(h,i)], effectively reducing their radius. Moreover, we observe a sudden drop of $\mathcal R$ for $\zeta=-3$ and positive $\al$. These correspond to the snapshots in Fig.~\ref{fig:snap}(o,n), which exhibit a highly regular organization of droplets with large sixfold symmetry.

To obtain more detailed insights into the spatial arrangement of droplets, we turn to the structure factor. To this end, we calculate the Fourier transform
\begin{equation}
    \hat\phi(\vec k,t) = \sum_i e^{\im\vec k\cdot\x_i}[\phi_i(t)-\phi_0]
\end{equation}
and with this the structure factor $S(k)=\mean{|\hat\phi(\vec k,t)|^2}$ binning the magnitude $k=|\vec k|$ of wave vectors $\vec k$. We extract the first moment
\begin{equation}
  k_\ast = \frac{\IInt{k}{0}{\infty} k S(k)}{\IInt{k}{0}{\infty} S(k)},
  \label{eq:kc}
\end{equation}
which yields the dominant length scale $\xi=2\pi/k_\ast$. Figure~\ref{fig:coex}(c) confirms that this length scale is indeed strongly correlated with the calculated droplet radius $\mathcal R$ as expected. We note that the outliers due to splitting (and merging) of a big droplet again fall below a master curve, which connects state points comprising many droplets (small $\xi$ and $\mathcal R$) with those exhibiting a single droplet (large $\xi$ and $\mathcal R$). The inset of Fig.~\ref{fig:coex}(c) shows the average number of droplets. This measure now clearly shows that some of the state points with $\gam<0$ still undergo full macroscopic phase separation and coarsen into a single droplet. In the regime of reverse Ostwald ripening, we observe a monotonic increase of the droplet number with $\alpha$.

\begin{figure}
    \centering
    \includegraphics{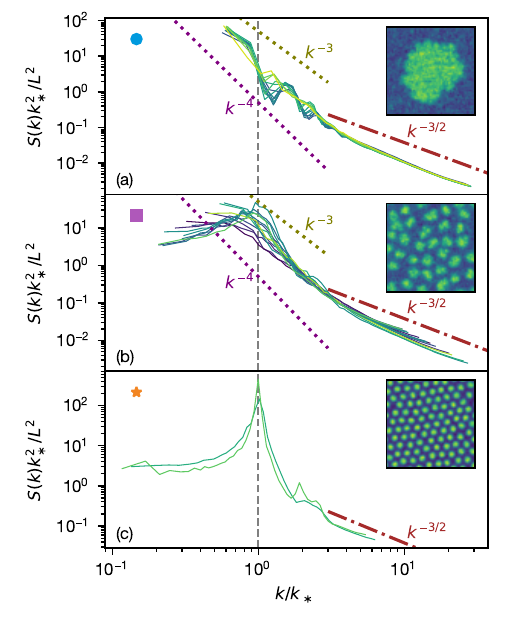}
    \caption{Structure factors $S(k)$ for (a)~the fully phase separated state with a single droplet, (b)~the disordered droplet phase, and (c)~hexagonally ordered droplets. Insets show example snapshots for each state. Quantities are normalized using $k_\ast$ defined in Eq.~\eqref{eq:kc} and the grid area $L^2$ ($L=128$). Dotted lines indicate power laws in the different regimes.}
    \label{fig:structfac}
\end{figure}

The full structure factors $S(k)$ are shown in Fig.~\ref{fig:structfac}, from which we find three qualitatively distinct cases. In the first case [Fig.~\ref{fig:structfac}(a)], $S(k)$ approaches $k\to0$ as a power law $k^{-\nu}$ with distinct regular oscillations. This behavior is indicative of phase-separated systems, which is confirmed by inspecting snapshots. Porod's law predicts the exponent $\nu=d+1=3$ in two dimensions~\cite{bray02}, which compares well with the numerical data. For large $k$, all structure factors seem to decay as $k^{-3/2}$. This scaling is due to discretizations errors in the finite difference stencil, which we have confirmed by decreasing the grid spacing leading to the expected $k^{-3}$ decay predicted through Porod's law (not shown). Importantly, the shape of $S(k)$ at smaller $k$ was not affected by changes in the grid spacing.

\begin{figure}[t]
    \centering
    \includegraphics{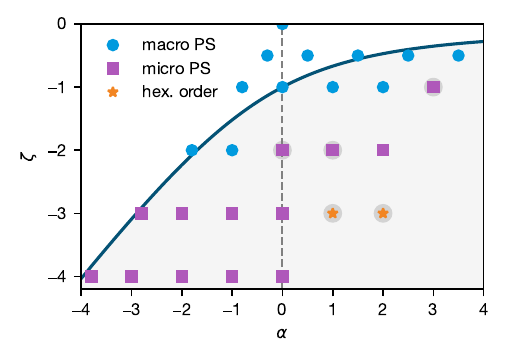}
    \caption{Transition from forward to reverse Ostwald ripening in the plane spanned by $\al$ and $\zeta$ under the influence of noise ($D=0.3$). The mean-field prediction for $\gam=0$ is shown as a solid line with $\gam<0$ inside the shaded area. We distinguish macrophase separation (i.e., a single droplet) from microphase separation (or droplet phase) and ordered droplets. Highlighted symbols correspond to outliers in Fig.~\ref{fig:coex}.}
    \label{fig:phasedia}
\end{figure}

The second class [Fig.~\ref{fig:structfac}(b)] is comprised of structure factors that cross over to a constant value as $k\to0$. As this limiting value becomes smaller, a broad peak develops that increases in height and moves towards $k_\ast$ without the distinct oscillations seen for a single droplet. Qualitatively, this peak indicates more and more irregular arranged droplets. For the two state points already noticed for their regular arrangement of droplets into a hexagonal lattice, the structure factor now exhibits a sharp peak at $k_\ast$ [Fig.~\ref{fig:structfac}(c)]. The other outliers noticed before (for $\zeta\in\{-1,-2\}$) are structurally closer to the droplet phase and characterized as those. Overall, the structure factor is a good descriptor to distinguish macrophase from microphase separation as shown in Fig.~\ref{fig:phasedia}, where we indicate the three different structural classes in the plane spanned by $\al$ and $\zeta$. In comparison with the mean-field prediction and the noiseless simulations [inset Fig.~\ref{fig:phi-alpha}(b)], in the presence of noise the droplet phase is shifted to larger values of $\al$ in particular at small $\zeta$.

\subsection{Droplet morphology}

\begin{figure}[t]
    \centering
    \includegraphics{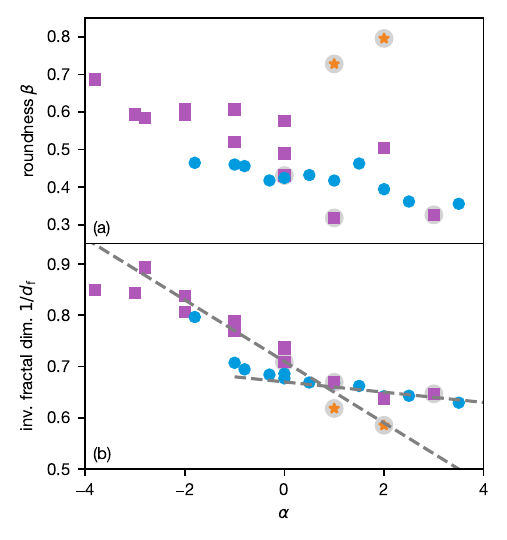}
    \caption{Droplet morphology. (a)~Average roundness $\beta$ and (b)~inverse of the fractal dimension $d_\text{f}$ plotted as a function of $\al$ for $D=0.3$ and $\phi_0=-0.4$. Dashed lines are guides to the eye. Same symbols as in Fig.~\ref{fig:phasedia}.}
    \label{fig:morph}
\end{figure}

In Fig.~\ref{fig:snap} we have seen that the droplets not only arrange differently in space but also exhibit distinct morphologies, which warrants further investigation. To this end, we introduce two measurements to quantify droplet morphology. First, we measure the area $A$ and perimeter $P$ of each droplet through counting pixels, defining the roundness
\begin{equation}
    \beta = \left\langle\frac{4\pi A}{P^2}\right\rangle
\end{equation}
so that $\beta=1$ for a perfectly circular droplet and any deviation leads to a reduction of $\beta$. We average over all sampled droplets. A pixel is defined as contributing to the perimeter of the droplet if it has one adjacent pixel that is not part of the droplet domain. The roundness as a function of $\al$ is shown in Fig.~\ref{fig:morph}(a). We observe that the roundness only weakly correlates with $\al$ and decreases as $\al$ becomes bigger. For macrophase separation the roundness is almost constant. The two state points that exhibit strongly ordered droplets are outliers showing the largest roundness.

Our second measure is the fractal Minkowski-Boulingand dimension
\begin{equation}
    d_\text{f} = \lim_{\varepsilon\rightarrow 0} \frac{\log(N(\varepsilon))}{\log(1/\varepsilon)}
\end{equation}
determined through box-counting with $\varepsilon$ the length of the boxes and $N(\varepsilon)$ the number of boxes necessary to cover the surface. In two dimensions, $d_\text{f} \in [1, 2]$ with $d_\text{f} = 1$ indicating a highly irregular and string-like object and $d_\text{f} = d = 2$ the smooth non-fractal limit. The results for $d_\text{f}$ are shown in Fig.~\ref{fig:morph}(b). We now observe a crossover from irregular shapes at negative $\al$ (in this regime droplets are thus more rounded according to $\beta$ but with a more fractal ``fuzzy'' boundary) with a large slope to a gentler slope for positive $\al$ indicating a smooth boundary. The largest $d_\text{f}$ again exhibit the ordered droplets, although we note that the droplets sampled at these two state points are only a few pixels in diameter and thus might bias the analysis.


\section{Conclusions}

We have studied different statistical measures for the spatial arrangement of droplets and their morphologies in Active Model B+ (AMB+), a minimal model for phase separation away from equilibrium. Compared to passive Model B, the current Eq.~\eqref{eq:j} now contains two non-potential terms with corresponding coefficients $\lam$ and $\zeta$. Importantly, while the $\lam$-term can be absorbed into an effective (pseudo) chemical potential [Eq.~\eqref{eq:mu}], the $\zeta$-term gives rise to a non-local contribution $\mu_\zeta$ [Eq.~\eqref{eq:delta}] that depends on the presence of interfaces but only affects curved interfaces.

We have performed extensive numerical simulations employing a GPU-based finite-difference code that efficiently handles the large stencils required for the $\zeta$-term~\cite{ambplusGitHub}. We have focused on steady state dynamics in a parameter regime where phase separation is expected ($a<0$) and dense droplets form ($\zeta<0$). Although noise has an impact and shifts transitions, we find that overall the mean-field predictions agree well with the numerical results. While $\zeta<0$ is a necessary condition, droplet properties are mainly governed by the single parameter $\al=(\zeta-2\lam)/\kap$ with the specific value of $\zeta$ playing a minor role. There are several robust trends: the number of droplets (in the droplet phase) increases with $\al$ [inset Fig.~\ref{fig:coex}(c)], the size of droplets remains approximately constant but droplets compactify. For negative $\al$ droplets are more fractal and become smoother as $\al$ increases. Intriguingly, inside the droplet phase we observe another transition, upon which droplets become more circular and order into an hexagonal lattice.

We also provide some additional insights into the mapping of non-potential terms onto an effective action through an integrating factor. Besides the Maxwell construction for droplets [cf. Fig.~\ref{fig:coex}], we have constructed the explicit effective action [Eq.~\eqref{eq:action}] and it would be interesting to derive reduced models for the dynamics of the droplet radius~\cite{fausti24,yan25}, which might also provide new approaches to the kinetics of droplet nucleation~\cite{cates23}. Moreover, flat interfaces in the mean-field theory are related through rescaling parameters by $\al$. Along $\al=0$, flat interfaces are given by Model B but droplets are still influenced by the non-local contribution due to $\zeta$.

Of course, the study of microphase separation and its microscopic origins has a long history in the theory of soft matter (emulsions, block copolymers, etc.)~\cite{leibler80}. On the level of a scalar order-parameter field (e.g., for the composition) the Landau–Brazovskii (LB) free-energy functional~\cite{brazovskii75} is widely used to model the transition to periodic patterns. We observe this ordering inside the droplet phase (see also Ref.~\cite{thomsen21}) and it would be interesting to understand if and in what limit the LB free energy is related to the effective action Eq.~\eqref{eq:action}.


\begin{acknowledgments}
  This work has been funded by the Deutsche Forschungsgemeinschaft (SFB 1551, Grant No. 464588647 and GRK 2516, Grant No. 405552959). Simulations have been performed through bwHPC on Helix. During the early stage of this project, JFR has been supported by the Alexander von Humboldt foundation. We gratefully acknowledge illuminating discussions with Sandra Schick and Samuel Shoup.
\end{acknowledgments}


\appendix

\section{Flat interface}
\label{appendix:flat}

For the calculation of $\gam$ [Eq.~\eqref{eq:gam}] we need to determine the profile $\phi^\infty$ of a flat interface. We first consider the case $\al=0$, for which $\rho=\phi$ reduces to the original field. In steady state it obeys the differential equation $\kap\partial_x^2\phi=a\phi+u\phi^3$ [cf. Eq.~\eqref{eq:mu}] with well-known solution ($w$ is the width of the interface)
\begin{equation}
  \phi^\infty(x) = \phi^\text{eq}_-\tanh(x/w), \qquad w = \sqrt{-2\kap/a}
\end{equation}
and thus
\begin{equation}
  \gam(\al=0) = \sqrt\frac{8(-a)^3\kap}{9u^2}\left(1+\frac{\zeta}{\kap}\sqrt\frac{-a}{u}\right).
\end{equation}
For $\al\neq 0$, we recast Eq.~\eqref{eq:mu} as
\begin{equation}
  \frac{\bar\mu^\infty}{\al} = \frac{a}{\al^2}\Phi + \frac{u}{\al^4}\Phi^3 - \kap\partial^2_{\tilde x}\Phi - \frac{1}{2}\kap|\partial_{\tilde x}\Phi|^2
  \label{eq:mu:scaled}
\end{equation}
rescaling length $\tilde x\equiv\al x$ so that the profile 
\begin{equation}
    \phi^\infty(x;\al,a,u,\kap) = \frac{1}{\al}\Phi(\al x;\tfrac{a}{\al^2},\tfrac{u}{\al^4},\kap)
\end{equation}
can be extracted from the scaling function $\Phi(\tilde x)$. It is easy to check that also $\psi$ is invariant under this scaling. Since we have already determined $\bar\mu^\infty$, we numerically solve Eq.~\eqref{eq:mu:scaled}.

\section{Stencils in finite-difference scheme}
\label{appendix:stencils}

Our direct numerical simulations deployed the discretization scheme of \textcite{tjhung18}, which used an Euler forward temporal stencil and a novel choice of spatial stencil to ensure stability with large time steps. The latter stencils are not fully described in Ref.~\cite{tjhung18} so we detail them here.

First-order derivatives are implemented via one of two stencils. The first are the ``normal'' $\order{\ell^8}$ central line stencils $\delta_x^{(8c)}$ and $\delta_y^{(8c)}$ approximating $\partial_x \phi$ and $\partial_y \phi$. The second is the block stencil
\begin{equation}
  \delta_x^{(\mathrm{TNC})} =
  \frac{1}{10\ell}\begin{bmatrix}
    -1 & 0 & 1 \\
    -3 & 0 & 3 \\
    -1 & 0 & 1
  \end{bmatrix}\,.
\end{equation}
The transpose of $\delta_x^{(\mathrm{TNC})}$ gives the derivative in the $y$-direction $\delta_y^{(\mathrm{TNC})}$. The Laplacian $\nabla^2 \phi$ is only calculated via the block stencil
\begin{equation}
  \Delta^{(\mathrm{TNC})} = \frac{1}{2\ell^2}\begin{bmatrix}
    -1 & 4 & -1 \\
    4 & -12 & 4 \\
    -1 & 4 & -1
  \end{bmatrix}\,.
\end{equation}
The Euler forward/It\=o integration step is
\begin{equation*}
    \begin{split}
        \phi(t + \Delta t) - \phi(t) =
        &- \left( \nabla \cdot \vec{j}^\mathrm{passive} + \nabla \cdot \vec{j}^\mathrm{active} \right) \Delta t
        \\ 
        &- \nabla \cdot \vec{j}^\mathrm{noise} \sqrt{\Delta t}
        + \order{\Delta t^2}\,,
    \end{split}
\end{equation*}
with discretised divergences
\begin{align*}
    \nabla \cdot \vec{j}^\mathrm{passive} &=
    \sum_k \delta_k^{(8c)} j_k^\mathrm{passive}\,, \\
    \nabla \cdot \vec{j}^\mathrm{active} &=
    \sum_k \delta_k^{(\mathrm{TNC})} j_k^\mathrm{active}\,, \\
    \nabla \cdot \vec{j}^\mathrm{noise} &=
    \sum_k \delta_k^{(8c)} j_k^\mathrm{noise}\,,
\end{align*}
where $k \in \{x, y\}$ and the deterministic currents are calculated via
\begin{align*}
    j_k^\mathrm{passive}
    &\approx -\delta_k^{(8c)} \mu^\mathrm{passive}\,, \\
    j_k^\mathrm{active}
    &\approx
    - \delta_k^{(\mathrm{TNC})} \mu^\mathrm{active}
    + \zeta \left( \Delta^{(\mathrm{TNC})} \phi \right) \delta_k^{(\mathrm{TNC})} \phi\,, \\
    \mu^\mathrm{passive} &\approx
    f'(\phi) - \kappa \Delta_k^{(\mathrm{TNC)}} \phi\,, \\
    \mu^\mathrm{active}
    &\approx \lambda \sum_k \left(\delta_k^{(\mathrm{TNC)}} \phi\right)^2\,.
\end{align*}
$j_k^\mathrm{noise}$ is Gaussian white noise with zero mean and variance $2 D \ell^{-2}$.

The block stencils $\delta_{\{x,y\}}^{(\mathrm{TNC})}$ and $\Delta^{(\mathrm{TNC})}$ are not the standard isotropic stencils used in similar problems (see e.g.\ Ref.\ \cite{patra06}). These non-standard stencils were chosen to enhances the numerical stability of solutions with non-integrable terms $\lambda \ne 0$ and $\zeta \ne 0$ \cite{tjhung18}. While each stencil involves 9-points ($\delta_x^{(8c)}$ and $\delta_y^{(8c)}$ are $1 \times 9$ and $9 \times 1$ line stencils), the $\delta_{\{x,y\}}^{(\mathrm{TNC})}$ and $\Delta^{(\mathrm{TNC})}$ block stencils are $\order{\ell^2}$ in error. The overall error of this scheme is therefore $\order{\ell^2}$.


%

\end{document}